\documentclass[twocolumn,showpacs,preprintnumbers,amsmath,amssymb,superscriptaddress,aps,prc]{revtex4}
\usepackage{graphicx}
\usepackage{xcolor}
\usepackage{subfigure}

\begin{document}

\title{Statistical features of the thermal neutron capture cross sections}

\author{M. S. Hussein}
\affiliation{Instituto de Estudos Avan\c{c}ados, Universidade de S\~{a}o Paulo C. P.
72012, 05508-970 S\~{a}o Paulo-SP, Brazil, and Instituto de F\'{\i}sica,
Universidade de S\~{a}o Paulo, C. P. 66318, 05314-970 S\~{a}o Paulo,
Brazil\\
Departamento de F\'{i}sica, Instituto Tecnol\'{o}gico de Aeron\'{a}utica, CTA, S\~{a}o Jos\'{e} dos Campos, S.P., Brazil}
\author{B. V. Carlson}
\affiliation{ Departamento de F\'{i}sica, Instituto Tecnol\'{o}gico de Aeron\'{a}utica, CTA, S\~{a}o Jos\'{e} dos Campos, S.P., Brazil}

\author{A. K. Kerman}
\affiliation{Physics Division, Oak Ridge National Laboratory, P.O. Box 2008, Oak Ridge, TN 37831, USA\\
Center for Theoretical Physics, Massachusetts Institute of Technology, 77 Massachusetts Avenue,  Cambridge, MA 02139, USA}

\begin{abstract}
We discuss the existence of huge thermal neutron capture cross sections in several nuclei. The values of the cross sections are several orders of magnitude bigger than expected at these very low energies. We lend support to the idea that this phenomenon is random in nature and is similar to what we have learned from the study of parity violation in the actinide region. The idea of statistical doorways is advanced as a unified concept in the delineation of large numbers in the nuclear world. The average number of maxima per unit mass, $<n_A>$ in the capture cross section is calculated and related to the underlying cross section correlation function and found to be $<n_A> = 3/(\pi \sqrt{2}\gamma_{A})$, where $\gamma_{A}$ is a characteristic mass correlation width which designates the degree of remnant coherence in the system. We trace this coherence to nucleosynthesis which produced the nuclei whose neutron capture cross sections are considered here.

\end{abstract}
\maketitle

\section{Introduction}

Very low energy neutron capture cross sections are important ingredients for nuclear research and applications. In the r-process of astrophysical significance, these cross sections are of fundamental importance as they dictate the path of nucleosynthesis. In applications, we mention energy production in reactors, and medical use in neutron capture therapy as well as in imaging. A recent compilation of these cross section is given in \cite{Mughab2003}. In Table 1, we show a sample of this compilation.

\begin{table}[htb]
\caption{Neutron capture cross section for several nuclei across the periodic table. The choice of the nuclei was dictated by the mass region and the disparity in the value of the thermal neutron capture cross section between adjacent nuclei or isotopes, when available. The full compilation can be found in \cite{Mughab2003}.
\label{table2}}
\begin{ruledtabular}
\begin{tabular}{lcc}
 &Nucleus & Cross section (barn) \\  
\hline
$Nucleus$ & & \\
\hline
&$^9{\rm Be}$ &[8.77$\pm 0.35]\text 10^{-3}$  \\
&$^{\rm10}{\rm B}$ & 0.5$\pm$0.0.1 \\
&n + $^{10}{\rm B}\rightarrow^{4}{\rm He} + ^{7}{\rm Li}$ & 3.8$\text x10^{3}$\\
&$^{14}{\rm N}$ & [79.8$\pm1.4]\text {x} 10^{-3}$ \\
&$^{15}{\rm N}$ & [0.024$\pm0.008]\text {x}10^{-3}$ \\
&$^{16}{\rm O}$ & [0.19$\pm0.019 ]\text{x} 10^{-3}$\\
&$^{20}{\rm Ne}$&[37$\pm4]\text{x}10^{-3}$\\
&$^{21}{\rm Ne}$&0.666$\pm0.110$\\
&$^{28}{\rm Si}$ & [177$\pm0.5]\text{x}10^{-3}$ \\
&$^{40}{\rm Ar}$ & 0.660$\pm$0.01 \\
&$^{40}{\rm Ca}$ & 0.41$\pm$ 0.02 \\
&$^{56}{\rm Fe}$ & 2.59 $\pm$0.14 \\
&$^{59}{\rm Co}$ & 37.18 $\pm$ 0.06 \\
&$^{58}{\rm Ni}$ & 4.5$\pm$0.2 \\
&$^{63}{\rm Cu}$ & 4.52$\pm$ 0.02 \\
&$^{84}{\rm Kr}$ & 0.111$\pm$0.015 \\
&$^{90}{\rm Zr}$ & 0.011$\pm$0.005 \\
&$^{103}{\rm Rh}$ & 145$\pm$2 \\
&$^{113}{\rm Cd}$ & 2.06 \text {x} $10^{4}\pm$400 \\
&$^{114}{\rm Cd}$ & 0.34$\pm$0.02 \\
&$^{149}{\rm Sm}$ & 4.014 \text {x}$10^{4}\pm$600 \\
&$^{157}{\rm Gd}$ & 2.54 \text {x} $10^{5}\pm$ 815 \\
&$^{159}{\rm Tb}$ & 23.3$\pm$0,4 \\
&$^{208}{\rm Pb}$ & 0.23$\pm$0.23 \\
&$^{209}{\rm Bi}$ & 0.0338$\pm$0.0007 \\
&$^{232}{\rm Th}$ & 7.35$\pm$0.03 \\
&$^{238}{\rm U}$ & 2.68$\pm$ 0.019\\

\end{tabular}
\end{ruledtabular}
\end{table}

It is a known fact that the thermal neutron (0.025 eV) capture cross section by $^{10}$B is about  0.5 barns. On the other hand the fission cross section of the reaction $n + ^{10}B \rightarrow ^{4}He + ^{7}{Li}$ is 3.8  x $10^{3}$ barns. Though the capture cross section for $^{10}$B is small, the absorption cross section is very large. We remind the reader that the absorption cross  section, intimately related to the strength function \cite{BF1963}, contains the capture cross section as a piece, plus other cross sections such as the above mentioned fission one.  In the case of heavy nuclei, one finds similar behavior. Take the case of Gadolinium 157. The thermal neutron capture cross section is about 2.54 x $10^{5}$ barns, to be contrasted with the capture by the other isotopes of Gadolinium, which are of much smaller value. In fact the capture by natural Gadolinium is 6 times smaller than that by $^{157}$Gd, yet it is still quite large owing principally to the presence of this isotope in the natural sample. As such the cross section for natural Gadolinium, extensively used as a contrast agent in Nuclear Magnetic Resonance (NMR) imaging, is

\begin{align}
&\sigma_{capture} = 0.0218 \sigma_{152} + 0.148 \sigma_{155} \nonumber\\
& + 0.2047 \sigma_{156} + 0.1565 \sigma_{157} + 0.2484 \sigma_{158}
\end{align}

For ultra cold neutrons ($E_n < $ 0.001 eV), the capture cross section for $^{157}$Gd can reach 1.2 x $10^{8}$ barns. This is comparable to typical atomic cross sections! The natural gadolinium capturecross section of these neutrons is about 4x $10^{5}$ barns. 
Other cases of large thermal neutron capture cross section are $^{153}$Cd, 2 x $10^{4}$ barns, and $^{135}$Xe, 3 x $10^{6}$ barns. The cadmium isotope $^{113}$Cd is commonly used as a neutron absorber-moderator in reactors and in other applications.\\
\begin{figure}[htb]
\begin{center}
\includegraphics[width=0.5\textwidth]{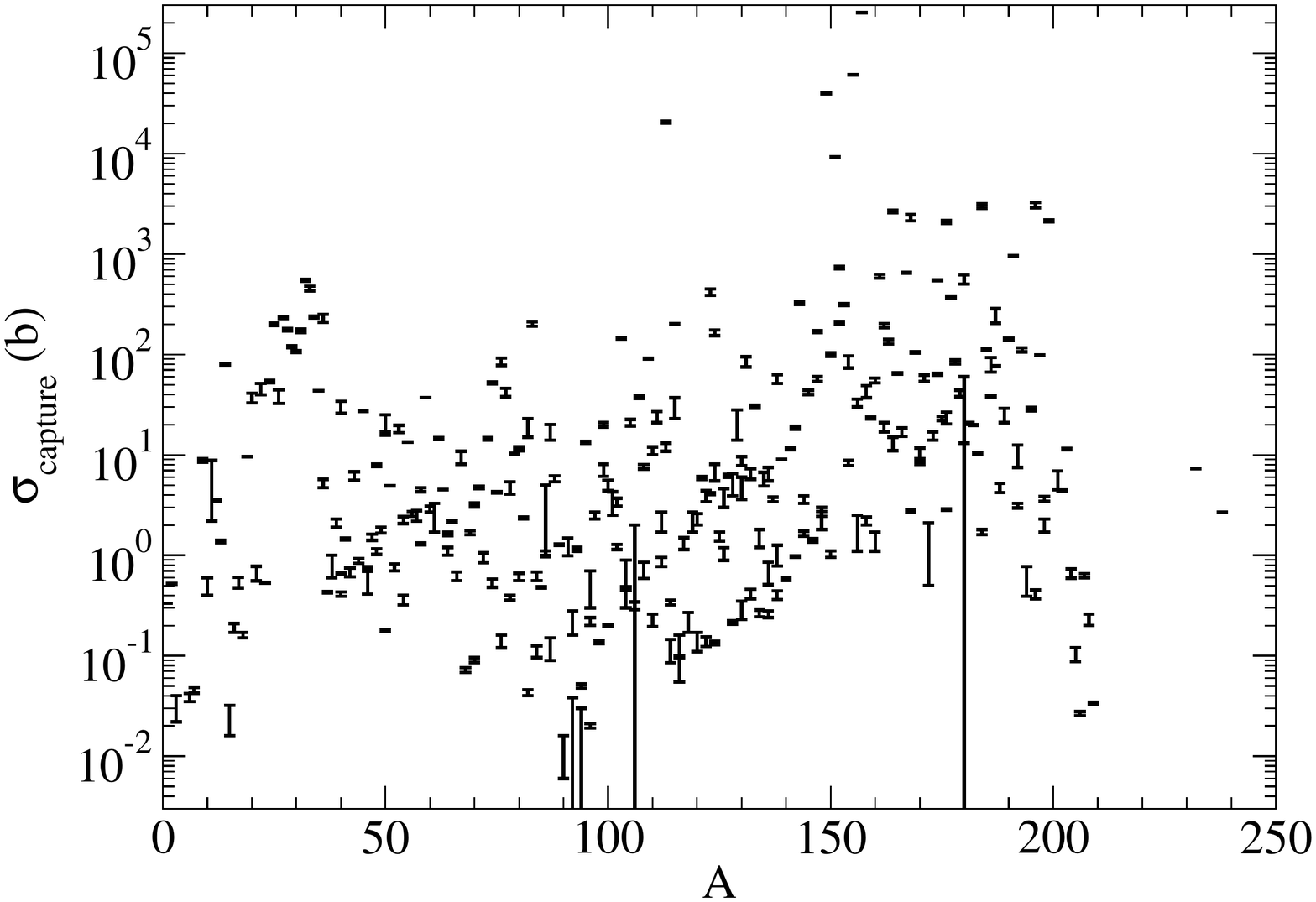}
\caption{Neutron capture cross sections vs. the mass number of the compound nuclei. The data were collected from the compilation of Ref. \cite{Mughab2003}.}
\label{fig1}
\end{center}
\end{figure}

In Figure 1 we show the thermal neutron capture cross section for a variety of nuclei. The Boron and Gadolinium cases stand out as notable exceptions of having an exceptionally large capture cross section. To be fair, the $^{10}$B capture cross section, (n, $\gamma$), is rather small. What is very large is the n-induced fission of the compound nucleus, $^{11}$B, namely, 

\begin{align}
&n + ^{10}B  \rightarrow  ^{11}B \rightarrow  ^{4}He (E_{\alpha} = 1.47 MeV) +  \nonumber \\   
&  + ^{7}Li (E_{Li} =0.84 MeV)  + \gamma (E_{\gamma} = 0.48 MeV) + 0. 231 MeV.
\end{align}

The reaction products, $\alpha$ and $^{7}$Li, are ionizing particles with a high linear energy transfer in environments such as living tissues,  and they lose all their energy within a  micrometer distance, roughly the diameter of the Boron-tagged cell.
The large values of the thermal neutron capture cross section in $^{157}$Gd; $\sigma_{157}$ =  2.54 x $10^{5}$; $^{153}$Cd; 2 x $10^{4}$ barns, and $^{135}$Xe, 3 x $10^{6}$ barns, and the  $^{10}$B($n_th$, $\gamma$)$\alpha$ + $^7$Li; $\sigma_{10}$ = 3.84 x$10^{3}$ barn reaction has received only minor attention as to their explanation.\\

In this contribution we take a critical look at the capture cross section data and present an analysis of both the fluctuating background using random matrix-inspired means, and the aforementioned anomalously large cases. In this latter case we base our discussion on the idea of a statistical doorway resonance which sits far up in energy but can influence the cross section in basically the same way that such doorways influence the parity violation ,"sign" problem", seen in the measurement of the longitudinal asymmetry of  epithermal neutrons scattered by thorium, uranium and other heavy nuclei \cite{bowman}. Other cases involving the statistical doorways r\"{o}le in resonance reactions, are the description of pre-equilibrium reactions \cite{FKK1980, BCHH1991}, and the decay of giant resonances \cite{DHA1986}. Of course, collective doorways, namely, states formed by coherent excitation of simple configurations, of 1p-1h coupled to 2p-2h states etc. are used in the description of giant resonance \cite{WS1972}.

\bigskip

\section{Abnormal nuclear resonance reactions and the possible r\"{o}le of simple doorways}

A notorious case of an abnormal resonance reaction, is the intermediate structure seen at low energy \cite{KRY1963}, and interpreted by Feshbach and Block \cite{BF1963} as arising from simple doorways, that modulates the compound nuclear resonances. We refer the reader to Feshbach's book on nuclear reactions \cite{Feshbach1993}. A more recent example, which has been already alluded to above, concerns the parity violation study using epithermal neutron scattering from several heavy nuclei. The results of the experiments, revealed a "sign" controversy, namely, by looking at the longitudinal asymmetry of the neutrons, A, it was found that the average $\langle A\rangle$  over the compound nucleus resonances, is predominantly positive, contrary to  one's belief that the average is zero in accordance with the statistical nature of the CN resonances. Several theoretical attempts were made to explain the "sign" problem \cite{bowmanreview, HKL95, FHKV00}. Quite recently, data on the distribution of reduced neutron widths of capture on Platinum were obtained and analyzed by \cite{koehler2010} and the idea was advanced that the usual, expected Porter-Thomas distribution breaks down. This finding prompted several theoretical works \cite{weiden2010, celardo2011}, which employ in one way or another a doorway mechanism to explain the deviation from the PT distribution. Conventional reaction theory without resorting to doorway was also attempted \cite{HK2011}. Related phenomena which may shed light on the resonance-dominated large capture cross section are the Stochastic Resonances \cite{WSB2004}, and Extreme Statistics \cite{LTBM2008}. In the following we discuss in detail the abnormally large thermal neutron capture cross section.

A compound nucleus resonance-dominated thermal neutron capture cross section can be written as,
\begin{equation}
\sigma_n = 4 \times10^{6} [barns]\frac{\Gamma_{n}\Gamma_{\gamma}}{(E - E_R)^{2} + 1/4(\Gamma)^2}
\end{equation}
The neutron width $\Gamma_n$ depends on energy and can be written as $\Gamma_n = \gamma_{n}\sqrt{E_{n}/1 eV}$, where $\gamma_n$ is the reduced width. At thermal energy, $E_n$ = 0.025 eV (T = 300 K, $v_n$ = 2200 m/s), the neutron width becomes about 0.1 meV, if a reduced width is taken as $\gamma_n = \overline{D}\times S_n$, where $S_n$ is the $l$=0 neutron strength function and $\overline{D}$ is the average spacing between compound resonances. From the systematics cited in \cite{BF1963} the strength function for an excitation energy of 8 MeV and A = 157, is about  5 $\times 10^{-4}$ . The $\gamma$ width $\Gamma_{\gamma}$ is about 0.15 eV.  For thermal energies, and A =157 (Gadolinium), $\overline{D}$ is  42.6 eV. Accordingly the ratio $\Gamma_{n}/\Gamma_{\gamma}$ = 6.8 $\times 10^{-4}$ is an extremely small number. For all practical purposes the total width in Eq. (3) is $\Gamma = \Gamma_{\gamma}$. Thus we can write for the $(n, \gamma)$ reaction, Eq. (3),

\begin{equation}
\sigma_n = 1.0 \times 10^{2}[barns\dot (eV)^2]\frac{1}{(E_n - E_R)^{2} + 1/4 (\Gamma_{\gamma})^2}
\end{equation}

If a resonance is close to the thermal energy, the above expression gives $\sigma_n$ = 1.78 $\times 10^{4}$ [barns]. The thermal neutron capture on $^{157}$Gd is $\sigma_n $= 2.26 $\times10^{5}$[barns]. However, the capture on the other stable isotopes of Gadolinium are orders of magnitude smaller (with the exception of $^{155}$Gd which has a capture cross section of $\sigma_n$ = 6.0 $\times 10^{4}$[barns]). The question that is asked is why the great variation in the value of the capture cross section. A resonance could be situated close to the thermal neutron energy in the case of $^{155}$Gd and in $^{157}$Gd, and not in the other isotopes. However, another estimate of the capture cross section can also be obtained for a resonance energy far from the thermal neutron energy, say, at $E_R$ = 22 eV. This gives $\sigma_n$= 2 barns, a huge difference from the $E_R$ = 0.025 eV case above. This difference of about 5 orders of magnitude, is what dictates the difference in the capture cross sections of the Gadolinium isotopes. But how accurate a measurement can be to be able to distinguish between an energy level in the compound nucleus at 8.0 + 2.5$\times 10^{-6}$ MeV from that at 8.0 MeV? This is hardly possible even with current advances in energy measurement techniques. The uncertainty in the position of the resonances in the compound nucleus prompted people to speculate that the occurrence of abnormal capture cross section is a random phenomenon. \\

The randomness idea can be better formulated using the concept of a doorway resonance sitting far away from the CN resonances, and having a total width much larger than that of the compound resonances $\Gamma_{D} >> \Gamma_{CN}$. Most of the  discussion to follow was invoked by Bloch and Feshbach back in 1963 in their seminal paper \cite{BF1963} on the fine structure seen in the neutron strength function $\langle \Gamma_{n} \rangle/\langle D \rangle$, below the usual giant structure. This intermediate structure was independentally introduced and discussed in \cite{KRY1963}. The doorway states are simple 2p - 1h states which are coupled to the neutron and $\gamma$ channels, and to the more complicated configurations in the compound system, 3p-2h, 4p-3h, etc. This latter coupling gives the doorway a spreading or damping width, $\Gamma^{\downarrow}_{D}$, the former accounts for the coupling to the open channels and gives the doorway an escape width, $\Gamma^{\uparrow}_{D}$. The door
 way states are also considered at higher energies in the so-called statistical multistep compound pre-equilibrium emission \cite{FKK1980, BCHH1991}. In these reactions the relative importance of the escape to the damping widths $\Gamma^{\uparrow}_{D^{i}}/ \Gamma^{\downarrow}_{D^{i}}$, of the different classes of the ever more complicated doorways populated in the reaction is very important. For a very recent review on compound nucleus reactions see \cite{CEH2014}. The important feature that distinguishes the doorway resonance from the CN resonance is that the total width of the doorway is $\Gamma_{D} = \Gamma^{\downarrow}_{D} + \Gamma^{\uparrow}_{D}$,  while that of the CN is just an "escape" width to the open channels. Using Feshbach's formula \cite{BF1963, FKL1967, Feshbach1993} , $\frac{\overline{\Gamma}_{CN}}{\overline{D}_{CN}} = \frac{\overline{\Gamma^{\uparrow}}_{D}}{\overline{D}_{D}}$, where D stands for the doorway resonance, we can estimate the average escape width
  
 of the doorway resonance (taken here to be a single isolated one).  The density of states of the 2p-1h doorway states is given by the formula, 

\begin{eqnarray}
\rho(E^{\star})_{2p-1h} & \equiv &  \frac{1}{\overline{D}_{2p-1h}} =  \frac{g(gE^{\star}-1/2)^{2}}{8 (2\pi)^{1/2}\sigma^{3}}  \nonumber\\
& & \times (2j +1)\exp{[-(j +1/2)^2/2\sigma^2]}
\end{eqnarray}
where $\sigma$ is the spin cutoff parameter, g is the average single particle level spacing near the Fermi level, given approximately by g = $\frac{A}{15}$, and the spin cutoff parameter is taken to be $\sigma^2 = 3\times 0.24A^{2/3}$. We take for the excitation energy, $E^{\star}$, the average neutron separation energy in the compound nucleus. We show in figure 2 a plot of the 2p-1h density vs. mass number.Taking for the excitation energy in the compound nucleus $^{158}$Gd, 8.0 MeV, we obtain for the density of 2p-1h states, the value (j = 2, 1), $\frac{1}{\overline{D}_{D}} = 20 MeV^{-1}$, see figure 2. This supplies the escape width of the doorway in $^{158}$Gd as $\Gamma^{\uparrow}_{D} = \Gamma_{\gamma}\frac{\overline{D}_{2p-1h}}{\overline{D}_{CN}} = 0.15 eV [\frac{50 keV }{42 eV}] = 0.18 keV$. An estimate of the damping width is more difficult to obtain. However, we can make a reasonable guess of $\Gamma^{\downarrow}_D \approx $1 keV. This will guarantee that the doorway will have an effect over $\Gamma^{\downarrow}_{D} / \overline{D_{CN}}$ = 1keV/ 42.6 eV = 22 CN resonances. With this value of the damping width, we can assess the  condition that the  doorway resonance is an isolated one in the sense, $\overline{\Gamma}_{D} / \overline{D}_{D}$ = [0.18[keV] + 1 [keV]] /50 [keV] = 0.024, a perfect condition for isolated resonances.

\begin{figure}[htb]
\begin{center}
\includegraphics[width=0.4\textwidth]{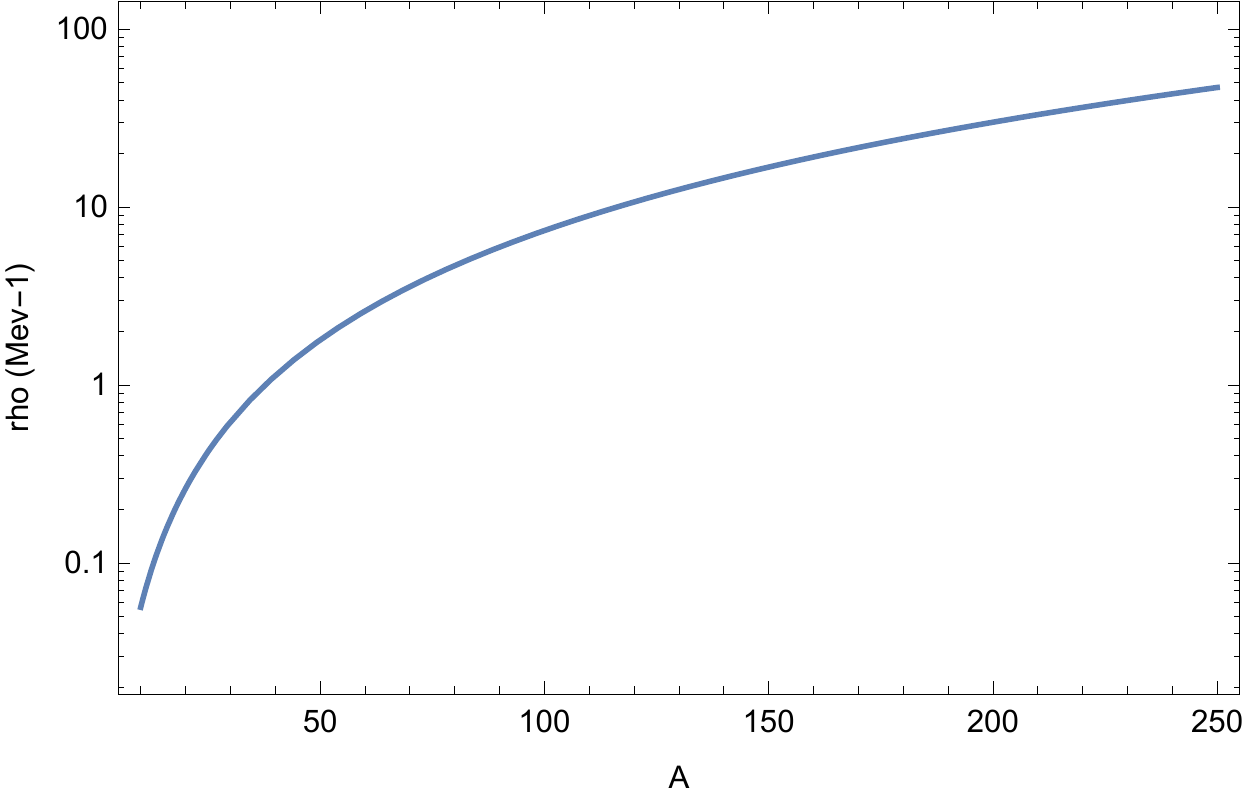}
\caption{ 2p-1h density of states vs. the mass number of the compound nucleus. The excitation energy, $E^{\star}$, for thermal neutron capture, $E_n = 0.025$ eV,  is taken to be the average neutron separation energy of the compound nucleus.}
\label{fig2}
\end{center}
\end{figure}

The coupling of the neutron channel to the 2p-1h doorway results in a modified expression for the capture cross section. To clearly demonstrate how the doorway resonance affects the capture cross section we resort to the relevant reaction theory as given by \cite{FKL1967, Feshbach1993}, and first write the amplitude for the $(n, \gamma)$ transition through the doorway as,

\begin{equation}
T_{n, \gamma} = \frac{g_{D, n} g_{D, \gamma}}{E - E_D + i\Gamma_{D}/2}
\end{equation}
where $g$ is the amplitude for the transition from the doorway to the open channel. The important difference between the doorway resonance and the compound resonance is that in the former, the total width of the doorway contains a damping width $\Gamma^{\downarrow}_{D}$, such that $\Gamma_{D} = \Gamma^{\downarrow}_{D} + \Gamma^{\uparrow}_{D}$, where the escape width, $\Gamma^{\uparrow}_{D}$  is the sum of all partial decay widths to the open channels, the usual width of a resonance. The compound nucleus width is basically an escape width. The partial width of the doorway is $\Gamma_{D, n} = 2\pi |g_{D, n}|^2$. We now make the assumption that the doorway state only couples to the neutron channel, and not to the $\gamma$ one. This means that the $\gamma$ emission proceeds from the compound nucleus resonances, $q$. Then, we can write $g_{D, \gamma} = \Gamma^{\downarrow}_{D}[E_D- E_q  + i\Gamma_{q}/2]^{-1}g_{q, \gamma}$, where $q$ labels the compound nucleus resonance. Of course experience has taught us that in general $\gamma$ emission can happen both from the doorway as well as from the compound nucleus. In fact, in the $\gamma$ decay of giant resonances in nuclei such as $^{208}$Pb, \cite{Beene1985,Dias1986} the two contributions are comparable. At the very low neutron energies considered here, and the excitation energies in the compound nucleus, sitting several MeV below the giant quadrupole resonances considered in \cite{Beene1985, Dias1986}, we ignore the "direct" doorway $\gamma$ decay, and consider this channel to be entirely open only to the compound nucleus resonances. Thus the cross section becomes, 

\begin{align}
&\sigma_n = \frac{1}{\pi^2} \times 10^{6} [barns] \frac{\Gamma_{D,n}|\Gamma^{\downarrow}_{D}|^2}{(E - E_D)^2 + 1/4(\Gamma_{D})^2}\times\nonumber\\
&\frac{\Gamma_{q,\gamma}}{(E_D - E_q)^2 + 1/4(\Gamma_q)^2}
\end{align}
The $q$ states are the compound nucleus resonances to which the doorway is coupled and we take them to be such that $E_q = E_D$, and $E_{D} \gg  E_n$, accordingly,
\begin{align}
&\sigma_n = \frac{1}{\pi^2} \times 10^{6}[barns]\frac{\Gamma_{D,n}\Gamma_{q,\gamma}(\Gamma^{\downarrow}_{D})^2}{(E_{D}^2 + 1/4(\Gamma_D)^2)(\Gamma_{q}/2)^2}\nonumber\\
& \approx \frac{4}{\pi^2}\times 10^{6}[barns][\Gamma_{D,n}]\frac{(\Gamma^{\downarrow}_{D})^2}{(E_D)^2 (\Gamma_q)}  \nonumber\\ 
& \approx \frac{1}{\pi^2}\times 10^{4}[barns/(eV)]\Gamma_{D,n} = 1.0 \times 10^{5} [barns]
\end{align}
where the neutron width of the doorway was in our model taken to be equal to its escape width of 0.18 keV. This is consistent with our assumption that the $\gamma$ decay proceeds only through the CN resonances in the vicinity of the doorway. The above cross section  is less than half in value of the empirical one 2.54$ \times 10^{5}$ [barns] cited above and listed in the compilation of \cite{Mughab2003}. In obtaining the above estimate we have used $E_D$ = 50 keV. Of course it is quite possible that the doorway could be located at a smaller energy. If we take $E_D$ = 30 keV, we would get for the cross section $\sigma_n = 2.7 \times 10^{5}[barns]$.\\

How frequent does such a doorway enhancement occur? We can estimate the probability of such a doorway enhancement by considering the ratio 
$\eta \equiv \Gamma_{D, n}/\Gamma_{q, n}$, which is the ratio of the cross section in the presence of the doorway, first equation in Eq. (8), to that without the doorway. The width $\Gamma_{q,n}$ is the usual CN neutron width when no doorway is present. What is random are the width amplitudes, $\sqrt{2\pi}g_{D,n}$ and $\sqrt{2\pi}g_{q,n}$, whose squares are the widths. At very low energies where the resonances are isolated these amplitudes are real. Call the distribution of the amplitudes $P(x)$.
The probability that the ratio $\eta = \frac{\Gamma_{D,n}}{\Gamma_{q,n}} = \frac{g_{D,n}^2}{g_{q,n}^2}$ defined above attains a certain value, $\eta_0$, is obtained by evaluating the integral,

\begin{equation}
P(\eta_0) = \int_{0}^{\infty} \int_{0}^{\infty} dx dy P(x)P(y)\delta(\frac{x}{y} - \sqrt{\eta_0})
\end{equation}

If a normalized Gaussian distribution is taken for P(x) and for P(y), the integral above can be readily evaluated to give,
\begin{equation}
P(\eta_0) = \frac{1}{2\pi}\frac{1}{1 + \eta_{0}},
\end{equation}
resulting in a very small probability for the occurrence of the doorway enhancement. Accordingly
very large values of neutron capture cross sections are inhibited by statistics.
\\

\begin{figure}[htb]
\begin{center}
\includegraphics[width=0.5\textwidth]{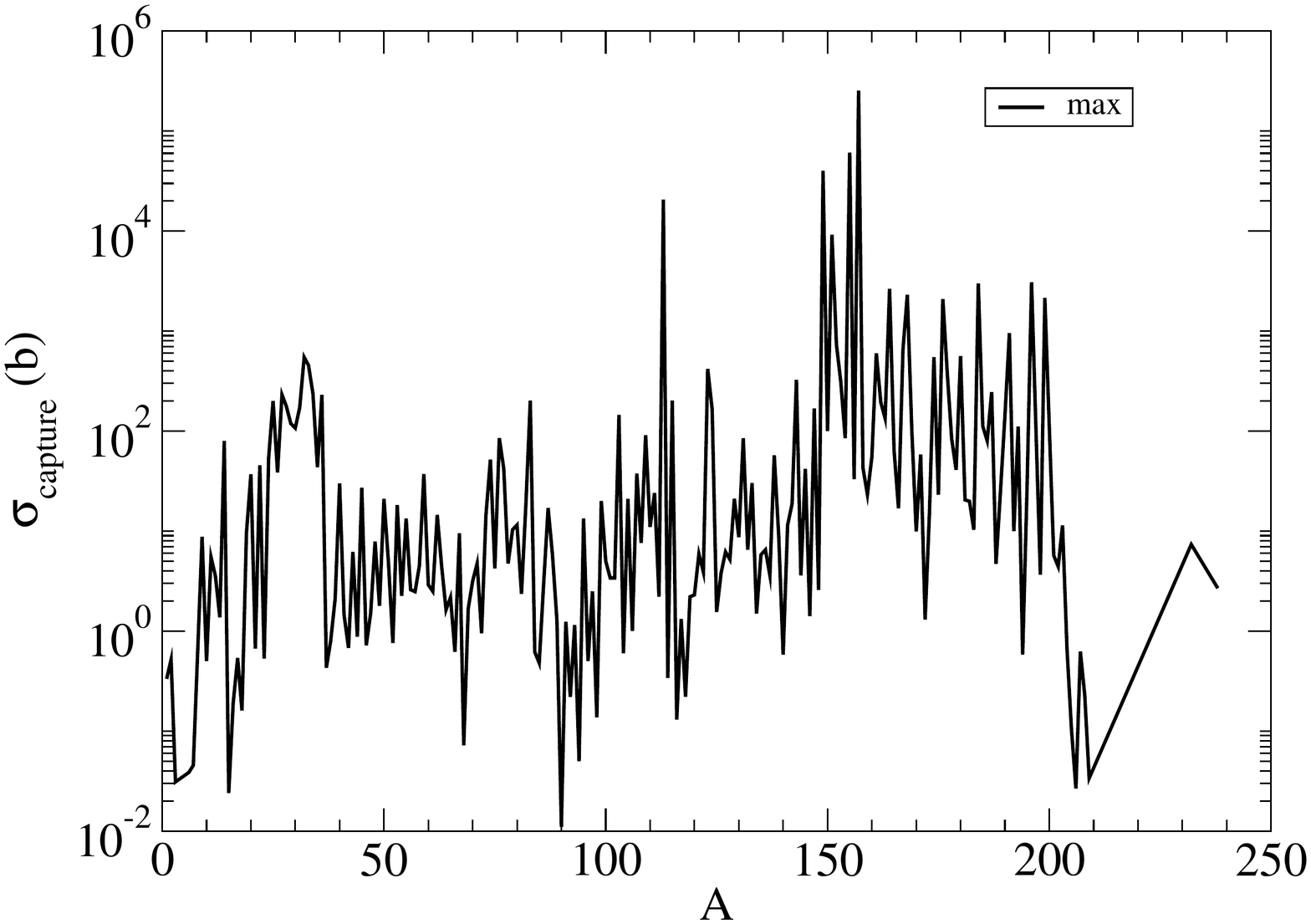}
\caption{Same as figure 1, but with a line passing through the fluctuating cross section to clearly exhibit the maxima.}
\label{fig3}
\end{center}
\end{figure}

\section{Average density of maxima in the capture cross section vs. A}
The thermal capture cross section vs. A is exhibited in Fig. 1. For the purpose of theoretical analysis to follow, we show in Fig. 3, the same as figure 1 but with a line that passes through the points. Further, in Figs. 4 -6 we present an enlarged figure 3, which exhibit the maxima in a clear and countable manner. One notices the abundance of fluctuations in $\sigma_{n}$ vs. A shown in these figures. These fluctuations may very well be random, though the capture cross section on a given nucleus as a function of the neutron energy is in the region of isolated resonances.\\

A measure of the statistical nature of the capture cross section which we propose here is the average number of maxima or minima in the cross section as a function of the mass number of the compound nucleus. This measure was suggested over 50 years ago by Brink and Stephen \cite{BS1963} for the cross section vs. bombarding energy, and it relies on Ericson's correlation function \cite{Ericson1963}. Later in  condensed matter theory, Efetov \cite{Efetov1995} worked out the correlation function in the case of variation of an external parameter such as an applied magnetic field on the shape of a nano devise such as an open quantum dot. He showed that the correlation function is the square of a Lorentzian, in contrast to Ericson's function for the variation with respect to energy, which is a Lorentzian. \\

Recently the results of Brink and Stephen were extended to Efetov's correlation function and subsequently to a general value of the tunneling probability, $p$, ranging between zero, for a closed system, to a maximum value of unity for an open quantum dot system \cite{RBHL2011, BHR2013, HR2014}. When applying Efetov's theory to nuclei, one would ask what is the external parameter? We trace the external parameter to the Universe which through Big Bang (BBN) and Stellar  nucleosynthesis created all the nuclei whose thermal neutron capture cross sections are shown in Fig.1.  For the purpose of theoretical analysis to follow, we show in Fig. 3, the same as figure 1 but with a line that passes through the points. Further, in Figs. 4 -6 we present an enlarged figure 3, which exhibit the maxima in a clear and countable manner. The correlation function, defined as $C(\delta A) \equiv [\langle \sigma(A) \sigma (A + \delta A)\rangle  - \langle \sigma(A)\rangle^2] / \langle \sigma(A)\rangle^2$ would  be $C(\delta A) = \frac{1}{[1+ (U(\delta A)/\overline{\gamma_{A}})^2]^2}$, where $U(\delta A)$ is the universal external parameter responsible for the creation of the nuclei shown in fig.1. We take this function to be linear in the variation $\delta A$, $U(\delta A) = c\delta A$ and accordingly define the correlation width $\gamma_A = \overline{\gamma_A}/c$. The Efetov correlation function is then, for a maximum value of the tunneling probability, $p$ = 1,

\begin{equation}
C(\delta A) = \frac{1}{[1 + (\delta A /\gamma_A)^2]^2}
\end{equation}
 \begin{figure}[htb]
\begin{center}
\includegraphics[width=0.5\textwidth]{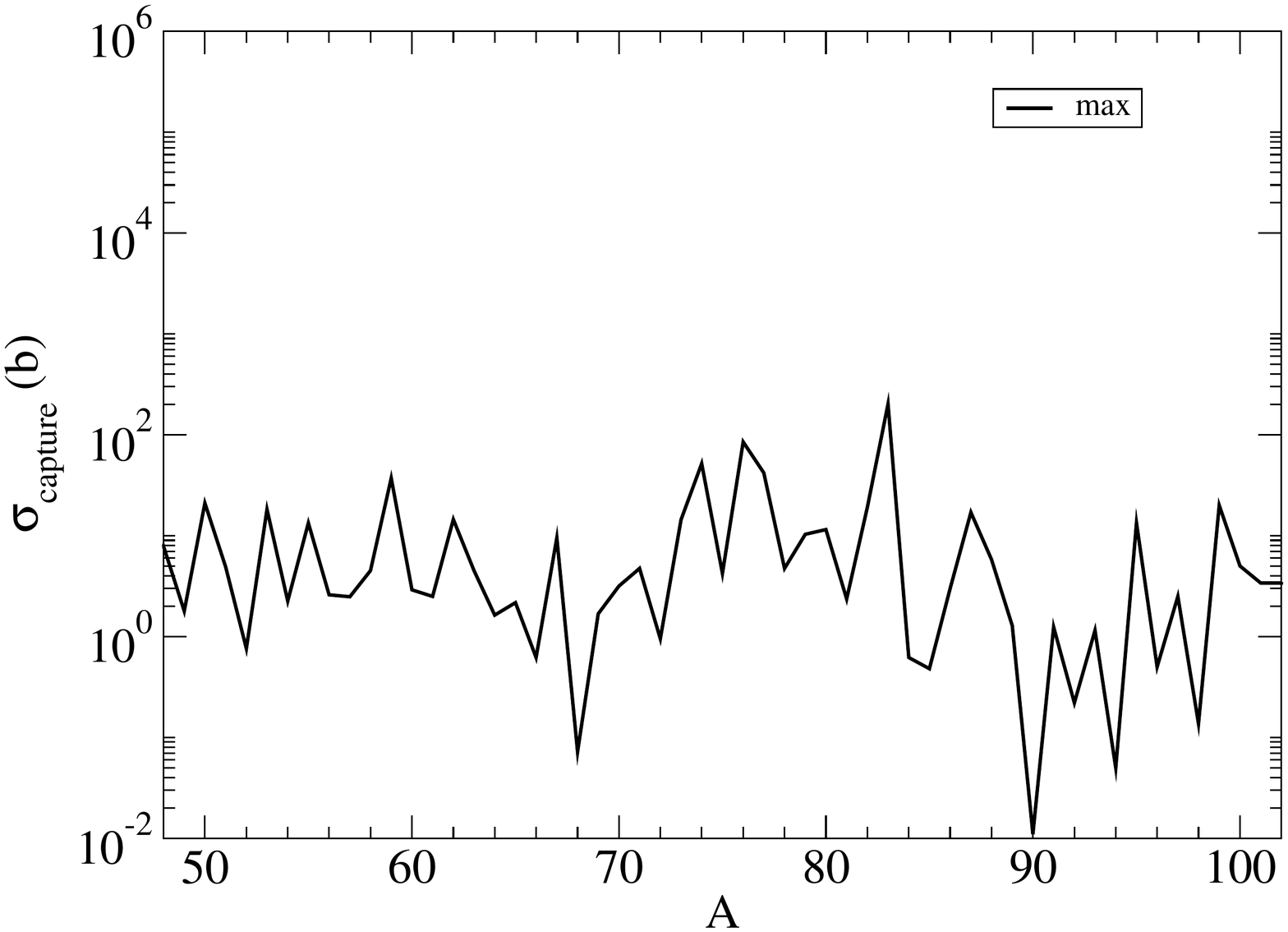}
\caption{Neutron capture cross sections vs. the mass number of the compound nuclei in the internal 50 $< $A $<$ 100 . The data were collected from the compilation of Ref. \cite{Mughab2003}.}
\label{fig1}
\end{center}
\end{figure}

\begin{figure}[htb]
\begin{center}
\includegraphics[width=0.5\textwidth]{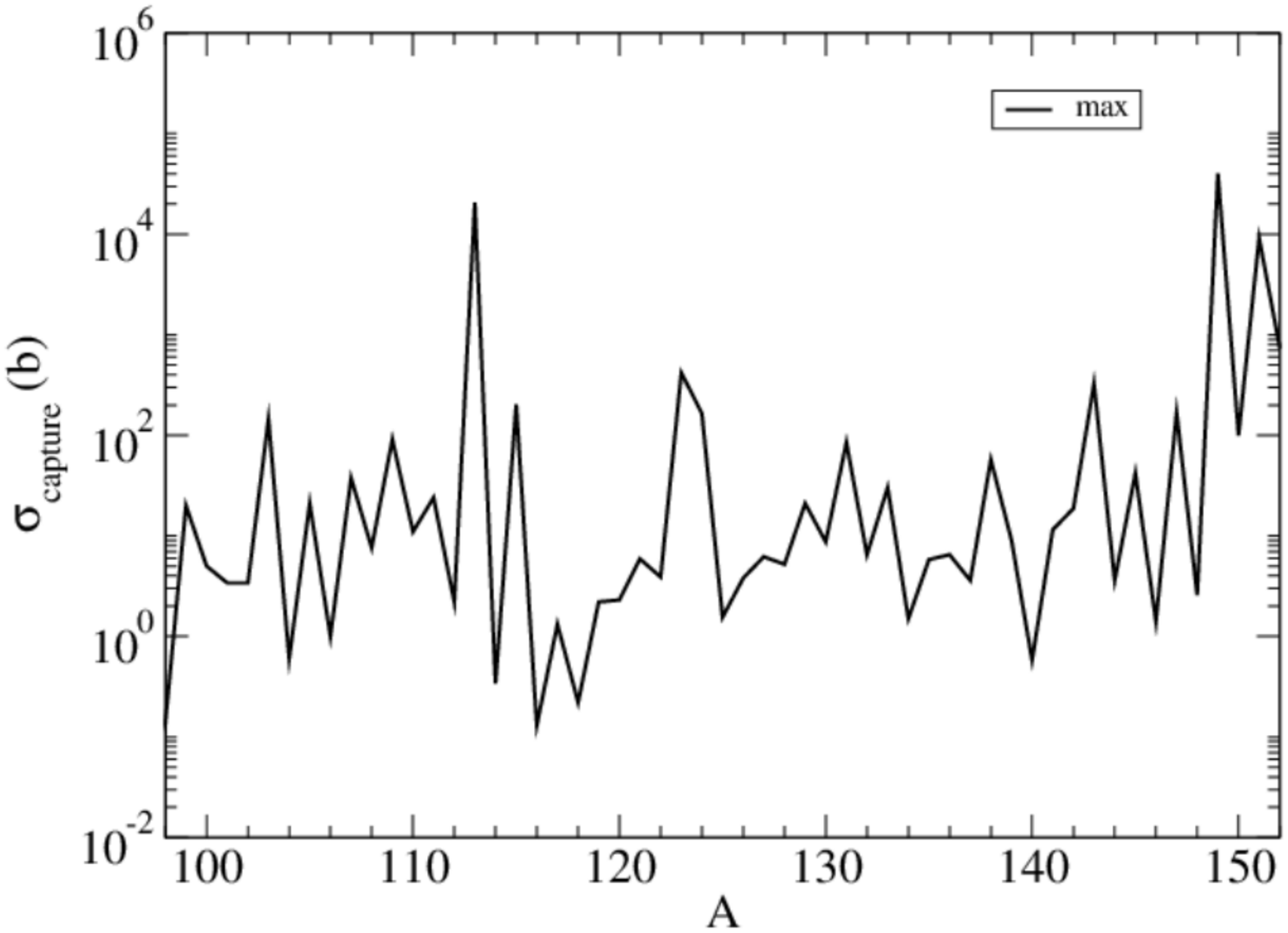}
\caption{Neutron capture cross sections vs. the mass number of the compound nuclei in the interval 100 $<$ A $< $150. The data were collected from the compilation of Ref. \cite{Mughab2003}.}
\label{fig1}
\end{center}
\end{figure}

\begin{figure}[htb]
\begin{center}
\includegraphics[width=0.5\textwidth]{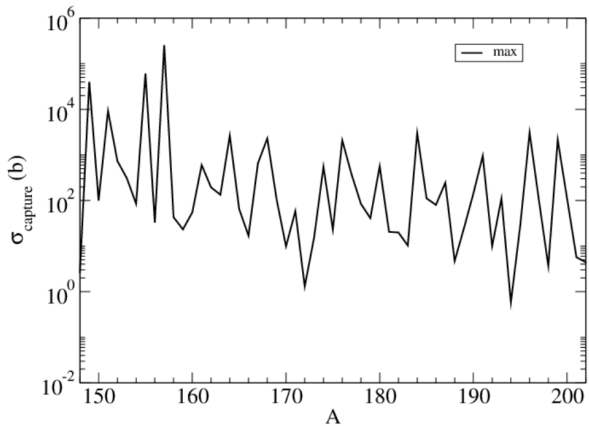}
\caption{Neutron capture cross sections vs. the mass number of the compound nuclei in the interval 150 $<$ A $<$ 200. The data were collected from the compilation of Ref. \cite{Mughab2003}.}
\label{fig1}
\end{center}
\end{figure}

Given a cross-section auto-correlation function, $C(z)$, the average density of maxima in the fluctuation cross section is found to be \cite{BS1963, RBHL2011},

\begin{equation}
\langle n_{z} \rangle = \frac{1}{2\pi}\sqrt{\frac{-C^{\prime\prime\prime\prime}(\delta z)\big|_{\delta z = 0}}{C^{\prime\prime}(\delta z)\big|_{\delta z = 0}}} \label{adm}
\end{equation}
Considering the general case of a tunneling or transmission probability in the interval $ 0 \le p \le 1$, the correlation function as a function of a variation in energy, $E$, or $A$ can be derived \cite{BHR2013},
\begin{equation}
C(\delta z) = \frac{A_{z}}{1 +(\delta z/\gamma_z)^2} + \frac{ B_{z}}{(1 + (\delta z/\gamma_z)^2)^2}\label{gcf}
\end{equation}
where, $A_{E} = 3p(2-p)-2$, $B_{E} = 4 + 4p (p-2)$, $A_{A} = 2p(1-p)$, and $B_{A} = 2+ p(3p - 4)$. The average density of maxima, Eq.~(\ref{adm}), is then given by, when the general correlation function of Eq.~(\ref{gcf}) is used, $\langle n_{z} \rangle = (\sqrt{3}/\pi \gamma_z)\sqrt{(A_z + 3B_z)/(A_z + 2B_z)}$

The tunneling probability alluded to above and used in the compound nucleus case, would be small in the  limit of weak absorption corresponding to  isolated resonances, $[\langle\Gamma_{q,n}\rangle/\langle D\rangle] \ll 1$, and unity in the case of strong absorption corresponding to  overlapping resonances, $[\langle\Gamma_{q,n}\rangle/\langle D\rangle]  \gg 1$. To turn these ratios into a probability we resort to the Moldauer-Simonius theorem \cite{Moldauer69,Simonious74} which states that in the general case the average S-matrix has the property, det$|\overline{S}| = e^{-\pi\Gamma/D}$ which in the one channel case gives $1 - |\overline{S}|^2 =1 - \exp{[-2\pi \langle\Gamma\rangle/\langle D\rangle]}$, where $\langle\Gamma\rangle$ is the average width of the compound nucleus. The tunneling probability is then taken to be an average transmission coefficient, $p =1 - |\overline{S}|^2$. 

Finally we can write for the average number of maxima in the cross section as the energy is varied $\langle n_{E} \rangle$ and as the mass number is varied $\langle n_{A}\rangle$ \cite{RBHL2011,BHR2013,HR2014},

\begin{equation}
\langle{n_{E}\rangle} = \frac{\sqrt{3}}{\pi \gamma_E}\sqrt{\frac{9p^2 - 18p + 10}{5p^2 - 10p + 6}}
\end{equation}
and
\begin{equation}
\langle{n_{A}\rangle} = \frac{\sqrt{3}}{\pi\sqrt{2} \gamma_{A}}\sqrt{\frac{7p^2 - 10p + 6}{2p^2 - 3p + 2}}
\end{equation}
where $\gamma_{E}$ is the correlation width of Ericson's fluctuations and $\gamma_{A}$ is the correlation width of Efetov fluctuations. In the limit of interest to us in the current paper, namely, $p < 1$, we  can set p = 0, and obtain,
\begin{equation}
\langle{n_{E}\rangle} = \frac{\sqrt{5}}{\pi \gamma_{E}}
\end{equation}
and
\begin{equation}
\langle{n_{A}\rangle} = \frac{3}{\pi \sqrt{2} \gamma_{A}}
\end{equation}

This last result is a new one in the nuclear context, and can be used directly to extract the correlation width $\gamma_{A}$ from the empirical data. In the case of compound nucleus fluctuations, we obtain for 3$\langle n_{A}\rangle$  = 18/50 + 23/50 + 17/50 = 1.16, see Figs. 4, 5, and 6.  Thus $\langle n_{A}\rangle$ = 0.39, and accordingly  giving for the correlation width, $\gamma_A$, the value 
\begin{equation}
\gamma_A = \frac{3}{0.39 \pi \sqrt{2} } = 1.94
\end{equation}

Accordingly, for all practical purposes, the remnant coherence  in the otherwise chaotic behavior of the capture cross section is restricted to $\Delta A$ = 1 and 2, which is expected as the nucleosynthesis which produced the nuclei occurs predominantly by adding one or two nucleons (s- and r-processes, notwithstanding BBN which involves several fusion reactions with $\Delta A >$ 2). The above findings also indicate the adequacy of using a fully statistical description of the compound nucleus, a known fact.  Of course the doorways are left out in this discussion as they correspond to extreme and rare events. \\
\section{Conclusions}
In conclusion, we have addressed the question of why the thermal neutron capture cross section by a very few nuclei is very large and escapes the normal trend found in most cases. We proposed that this effect may be traced to simple 2p-1h doorway states that accidentally affect the neutron capture in some nuclei. The chance for this to happen is very small as required by the data. We have also suggested a new measure of the degree of chaoticity of the compound nucleus cross sections based on the average density of maxima. Our findings could potentially be of value in finding other cases of very large capture cross sections and possible application to the study of radiative capture involving exotic nuclei, of relevance to the s-process in astrophysics. \\

{Partial support from the CNPq and FAPESP are acknowledged by BVC and MSH.}\\

\end{document}